\documentclass[10pt]{article}
\include{epsf}
\begin{document}

\title{\large \bf Semiclassical Quantization and Resonance\\
in Spin Tunneling\footnotemark[2]\footnotetext[2]{Dedicated to Heinz
Horner on the occasion of his 60th birthday.}}

\author{J.L. van Hemmen \\
Physik-Department der TU M\"{u}nchen \\
D-85747 Garching bei M\"{u}nchen \\ Germany
\\ \\ and \\ \\  
A. S\"{u}t\H{o} \\
Research Institute for Solid State Physics \\ P.O. Box 49,
H-1525 Budapest 114 \\ Hungary}

\date{}

\maketitle

\thispagestyle{empty}

\begin{abstract}

\noindent We derive semiclassical quantization for a spin, study it 
for not too small a spin quantum number ($S \ge 5$), and compute the 
$2S+1$ eigenvalues of a Hamiltonian exhibiting resonant tunneling as
the magnetic field parallel to the anisotropy axis is increased.  
Special attention is paid to the resonance condition.  As a corollary 
we prove that semiclassical quantization and quantum-mechanical 
perturbation theory agree there where they should.

\end{abstract}

\vspace{1cm}
\vspace{1cm}

\newpage

\section{Introduction}

Resonance has revived the experimental interest in spin quantum
tunneling. The idea is both fascinating and simple
\cite{NoSe}--\cite{Mnd}. Each molecule in, e.g., a ${\rm Mn}_{12}$
acetate crystal carries a spin of fixed angular momentum $10 \hbar$ and 
experiences a constant magnetic field {\bf H}.  The corresponding 
Hamiltonian is given by
\begin{equation}
	{\cal H} = -\gamma S_z^2 -g \mu_{\rm B} {\bf H}\cdot{\bf S}\ .
	\label{1}
\end{equation}
The first term on the right is a magnetic anisotropy.  Now let ${\bf
H} = (\alpha, 0, \delta) $ and absorb $g \mu_{\rm B} $ into $\alpha$
and $\delta$ so that Eq.~(\ref{1}) reduces to
\begin{equation}
	{\cal H} = -[\gamma S_z^2 + \delta S_{z}] - \alpha S_{x}
	\equiv - F(S_{z}) - \alpha S_{x} \ .
	\label{2}
\end{equation}
Here $\gamma>0$ and we may assume $\alpha,\,\delta\geq 0$.  By putting 
$\gamma:=-\gamma$ we encounter a situation whose tunneling physics and 
mathematics hardly changes and therefore need not be treated 
separately.  In the present paper, $S_{x}$ and $S_{z}$ have the 
dimension of angular momentum, to be measured in units $\hbar$.  
For the moment we suppose that the spin quantum number $S$ is an
integer.

The term $\alpha S_{x}$ tries to generate a rotation about the 
$x$-axis and thus aims at inducing a tunneling transition.  It 
certainly does so for $\delta =0$.  For arbitrary nonzero $\delta$ the 
degeneracy in the spectrum of $S_{z}^2$ is lifted and no tunneling can 
occur \emph{unless} a special choice of $\delta$ is made that restores 
the degeneracy.  For $\alpha=0$ a degeneracy exists, if $F(m \hbar) = 
F(n \hbar)$ for some integers $m$ and $n$, i.e., putting $\gamma \hbar 
= \Gamma$,
\begin{equation}
	\Omega_{m}^{(0)} \equiv -m^{2} \Gamma -m \delta = -n^{2} 
	\Gamma -n\delta \equiv \Omega_{n}^{(0)}\ .
	\label{3}
\end{equation}
In passing we note that both $\Gamma$ and $\delta$ have the dimension
of frequency. Equation (\ref{3}) tells us
\begin{equation}
	(m^{2} - n^{2}) \Gamma + (m-n) \delta = 0
	\label{4}
\end{equation}
so that with $-S \le m \neq n \le S$ we are left with
\begin{equation}
	(m+n) \Gamma + \delta = 0 \Rightarrow  m+n = -\delta /\Gamma =-k
	\label{5}
\end{equation}
where $k \ge 1$ is chosen to be a positive integer.  Hence $m = -n-k$.  
The remarkable aspect of this condition is that it can be realized for 
\emph{all} $n$ in the ballpark $-S \le n \le S-k$.  Furthermore, 
degeneracy can be realized by a single choice, viz., $\delta = k 
\Gamma$.  Here one has exploited the fact that the anisotropy is 
\emph{quadratic} in $S_{z}$.  The variable parameter $\delta$ is at 
the disposal of the experimentalist and the corresponding magnetic 
field is usually increased from minus to plus a few Tesla.  One then 
finds \cite{NoSe}--\cite{Mnd} several resonances in between.  It is
fair to say that the above setup is indeed elegant.  It is a corollary 
of semiclassical quantization as treated in the present paper that the 
resonance condition (\ref{5}) also holds for $\alpha \neq 0$.  The 
(tiny) level splitting is a consequence of quantum mechanics.

Meanwhile resonance has been studied intensively.  Already those who 
invented it initiated a first attempt \cite{Mna} at its theoretical 
understanding.  In an admirable \emph{tour de force}, Chudnovsky and 
Garanin \cite{cg,gc} were able to fully analyze the influence of the 
heat bath provided by the surroundings of the spins on the tunneling 
process if $\delta = 0$, and there is little doubt that their 
arguments can be generalized to the $\delta \neq 0$ case.  Here we 
assume a much lower temperature so that the heat bath can be 
neglected.  The analysis of semiclassical quantization, including 
resonance, is the main theme of the present paper. 
It yields the energy eigenvalues of the Hamiltonian (\ref{2})
to a good approximation and thus completes our earlier work
\cite{vhs4,vhs5,vhs6} on the WKB formalism for spins; we refer the 
reader to the excellent review \cite{Gunther} for additional 
information regarding the experimental and theoretical context.  
In \S2 we reconsider
the WKB formalism and indicate the use of the WKB wave functions in 
computing the level splitting at resonance.
We then treat semiclassical spin quantization (\S3) and determine (\S4)
the 
energy eigenvalues for arbitrary external field, i.e., $\delta$. In
\S5 we extend the results to half-integer spin quantum numbers 
and Hamiltonians in 
which the term $\alpha S_x$ is replaced by $\alpha S_x^l$ with $l>1$. 
The conclusion is that semiclassical quantization holds here too
but that for even $l$ the condition (\ref{3}) may correspond to true 
degeneracy; that is, no resonance and no tunneling. 
We finish the paper with a discussion (\S6).

\section{WKB and Resonance}

The eigenvalues and eigenvectors of the Hamiltonian (\ref{2}) are 
solutions of the eigenvalue equation ${\cal H}\psi=E\psi$.  The
$S_{z}$ axis being the main anisotropy axis, it is natural to write 
this equation in a representation with $S_{z}$ diagonal.  Then $S_{z}$ 
is simply $s=n\hbar$, a multiplication operator on the spectrum of 
$S_{z}$.  Furthermore, let $T_{\pm \hbar}$ induce a translation by 
$\pm \hbar$ so that $(T_{\pm \hbar} \psi)(s) = \psi (s \pm \hbar)$.  
Then $S_{x}$ reads
\begin{equation}
S_x = \frac{1}{2} \, (S_+ + S_-) = \frac{1}{2} \left[ a \left(
       \sqrt{s(s+\hbar)} \, \right) T_\hbar + a \left(
	\sqrt{s(s-\hbar)} \, \right) T_{-\hbar} \right]
\label{6}
\end{equation}
where $a(s) = [ \sigma(\sigma + \hbar) - s^2 ]^{1/2}$.  
As a consequence, the Schr\"odinger equation assumes the form of a 
second-order difference equation,
\begin{equation}\label{Sch}
h_{n,n-1}\psi_{n-1}+h_{n,n+1}\psi_{n+1}+(h_{nn}-E)\psi_n=0 \ .
\end{equation}
The matrix elements and the vector components are taken in the 
basis of the eigenvectors of $S_z$, and (\ref{Sch})
is valid for $-S+1\leq n\leq S-1$. Disregarding the remaining
two equations, relation (\ref{Sch}) has two linearly
independent solutions for {\em any} value of $E$; they are determined
by fixing, e.g., $\psi_0$ and $\psi_1$. The $2S+1$ 
eigenvalues of $\cal H$ are singled out by requiring that 
$\psi$ satisfy the boundary conditions
\begin{equation}\label{bc}
h_{\pm S,\pm(S-1)}\psi_{\pm(S-1)}+(h_{\pm S,\pm S}-E)\psi_{\pm S}=0\ .
\end{equation}
Equivalently, we can extend (\ref{Sch}) to $n=\pm S$ by defining 
arbitrary real $h_{\pm S,\pm (S+1)}= h_{\pm (S+1),\pm S}$ and imposing 
the boundary conditions $\psi_{\pm(S+1)}=0$.

If $\alpha=0$, the eigenvectors are those of $S_z$ and the eigenvalues 
equal $\hbar\Omega_n^{(0)}$ with $n$ ranging from $-S$ to $S$; cf.  
Eq.~(\ref{3}).  It is interesting to note that, as the energy 
increases while avoiding accidental degeneracies (\ref{3}), the 
eigenvectors are alternately localized on the `right' ($n>0$) or on 
the `left' ($n<0$).  Far from degeneracy an approximate localization 
on alternating sides remains true for a nonvanishing $\alpha$, if 
$\alpha\ll\Gamma S+\delta$, which is supposed throughout the paper.

By varying $\delta$, neighbouring eigenvalues can get very
close to each other, although true degeneracy cannot
occur \cite{nw}. As we discuss below, the mechanism of the avoided 
level crossing is spin tunneling. When two eigenvalues become as close
as possible,
we speak of a {\em quantum-mechanical resonance}. This has to
be distinguished from what we call a {\em semiclassical resonance},
which we understand to be the coincidence of semiclassical eigenvalues.
We shall prove in \S4 that the condition for a semiclassical resonance 
remains (\ref{3}). Apart from the resonance at $\delta=0$, we cannot 
expect that the two definitions predict the same `resonant'
values for $\delta$. Because, however, semiclassical
eigenvalues nicely approximate the true ones, only a small `fine tuning'
of $\delta$ may be necessary to pass from semiclassical to true,
quantum-mechanical, 
resonance. In the remaining part of this section we explain how one can
estimate the level splitting at resonance.

Let us choose $\delta$ close to a resonant value and let $E_1<E_2$ be 
two (unknown) neighbouring, nearly degenerate, eigenvalues.  Let us 
also imagine that we are given two linearly independent vectors 
$\eta(E)_n$ and $\vartheta(E)_n$ which depend continuously on $E$ and 
solve (\ref{Sch}).  To avoid all confusion, we emphasize that no 
linear combination of them satisfies the boundary conditions 
(\ref{bc}) and, hence, the eigenvalue equation, if $E$ is not an 
eigenvalue.  On the other hand, the eigenvector belonging to $E_1$ is 
a linear combination of $\eta(E_1)$ and $\vartheta(E_1)$, and its 
analog holds for $E_2$.  Let $E_1^{\rm (sc)}$ and $E_2^{\rm (sc)}$ be 
the semiclassical eigenvalues corresponding to $E_1$ and $E_2$, 
respectively.  Because all four energies are now close to each
other, we can replace $\eta(E_1)$ and $\eta(E_2)$ by $\eta(E_1^{\rm 
(sc)})$, and $\vartheta(E_1)$ and $\vartheta(E_2)$ by 
$\vartheta(E_2^{\rm (sc)})$, or {\em vice versa}.  In this way, 
finding $E_1$ and $E_2$ {\em and} the corresponding eigenvectors 
reduces to good approximation to diagonalizing $\cal H$ in a 
two-dimensional subspace.  In so doing, we can use the WKB method to
obtain $\eta(E)$ and $\vartheta(E)$.

Though Wentzel, Kramers, and Brillouin (WKB) never considered spins, 
their idea is applicable here too, provided one generalizes the 
formalism appropriately \cite{vhs4,vhs5,vhs6} so as to take care of 
the discrete nature of a spin and its different commutation relations 
as compared to a particle.  A semiclassical analysis formally means 
that we take the limit $\hbar \to 0$ and at the same time $S \to 
\infty$ in such a way that $\hbar S = \sigma$ remains constant.  In 
this limit we are left with a continuum description with $s$ ranging 
in the interval $[-\sigma, \sigma]$.  In the spirit of WKB we now make 
the ansatz $\psi = \exp (i {\cal S}/\hbar)$ with 
\begin{equation}
	{\cal S} = {\cal S}_{0} + \sum_{n=1}^{\infty} ({\hbar \over i})^{n} 
	{\cal S}_{n}
	\label{7}
\end{equation}
for the wave function we are looking for,
expand everything in powers of $\hbar$, and usually stop after the 
first-order term, the zeroth-order one being dominant.  So
\emph{de facto} we use a continuum description, even though $\hbar$ is
still finite.

\begin{figure}
 \epsfxsize=10 truecm
 \centerline{\epsffile{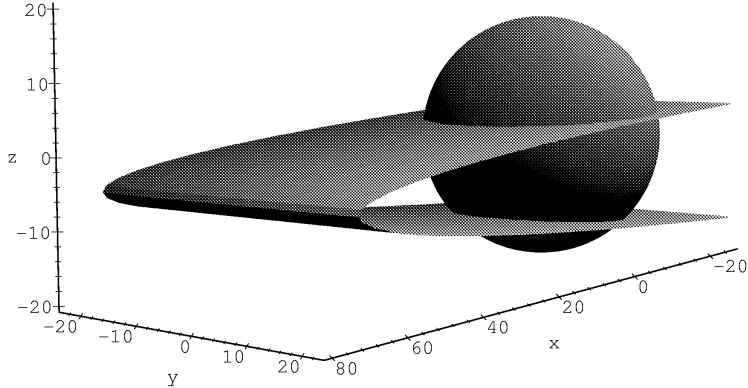}}
 \caption{Classical orbits of the Hamiltonian ${\cal H} = -\gamma S_z^2-\delta 
 S_z -\alpha S_x$ are the intersection(s) of the energy surface 
 $E=-\gamma S_z^2-\delta S_z-\alpha S_x$ and the sphere 
 $S_x^2+S_y^2+S_z^2=\sigma^2$, here plotted for $S=20,\ 
 \hbar=\alpha=\gamma=1,\ \delta=10$, and $E=-50$.}
\label{fig1}
\end{figure}   

The dominant contribution to tunneling comes from the classically 
forbidden region, say, between the inner turning points $b_{1} < 
b_{2}$.  In this interval the two linearly independent WKB wave 
functions read \cite{vhs4,vhs5,vhs6}
\begin{equation}
	\phi_{E, l} (s) =  C_l \exp -{1 \over
        \hbar}\int_{b_{1}}^{s} 
	{\rm d} s' {\rm arccosh} \left ({-E -F(s') \over \alpha a(s')}
	\right)
	\label{wkbl}
\end{equation}
and
\begin{equation}\label{wkbr}
	\phi_{E, r} (s) =  C_r \exp -{1 \over
\hbar}\int_{s}^{b_2} 
	{\rm d} s' {\rm arccosh} \left ({-E -F(s') \over \alpha a(s')}
	\right)\ .
\end{equation}
Here $C_{l,r}$ are normalization constants and the arccosh
expression stems from ${\cal S}_{0}$ in
(\ref{7}).  As can be seen from Figs.\ 1 and 2, there are four turning 
points on the $s:= S_{z}$ axis, two inner ones, $b_{1}$ and $b_{2}$, 
and two outer ones, $a_{1}$ and $a_{2}$.  The inner turning points 
$b_{1}$ and $b_{2}$ are boundaries of the classically allowed motion 
to the left and right of them, respectively, and such that here the 
argument of the hyperbolic cosine equals 1.  The functions depend 
continuously on the (classical) energy $E$. By using appropriate
connection formulae, both can be extended to the whole interval
$[-\sigma,\sigma]$, and the functions obtained in this way are almost
perfectly localized on the left and on the right, respectively.  
Indeed, between $b_1$ and $b_2$ Eqs.\ (\ref{wkbl}) and (\ref{wkbr})
define, respectively, a rapidly decaying and a rapidly increasing 
function.  Localization of the extended solutions then follows from 
$\phi_{E,l}(b_1)\gg\phi_{E,l}(b_2)$ and 
$\phi_{E,r}(b_1)\ll\phi_{E,r}(b_2)$.  For later use we note that the 
two functions are not orthogonal to each other and, because of the 
localization, their tiny overlap comes essentially from the 
classically forbidden region.

\begin{figure}
 \epsfxsize=9 truecm
 \centerline{\epsffile{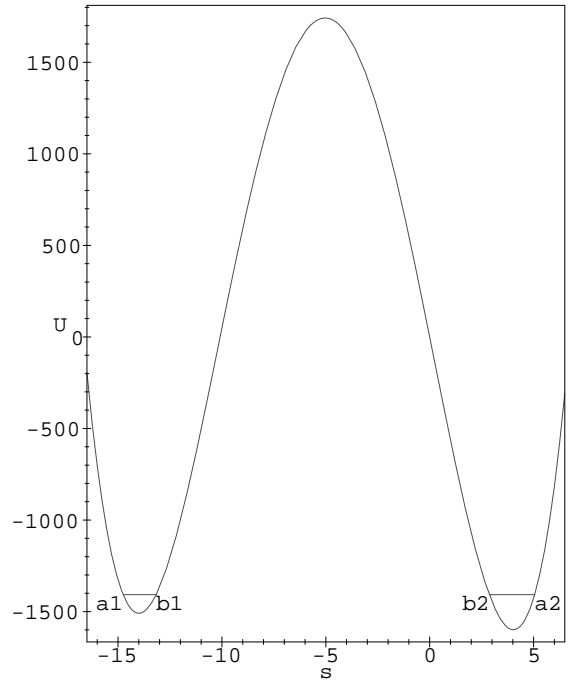}}
 \caption{
The effective, quartic double-well potential $U(s)$ given by 
Eq.~(\ref{9c}) has been plotted for $\hbar=\alpha=\gamma=1$, 
$\delta=10$, and $E=-56.7$. For $S=20$, $E=E_{-14}=E_4$; that is,
the levels -14 and 4 are at resonance. The two horizontal segments
are located at the `energy' level ${\cal E}=(S^2-E^2)/2$. 
They end up in the turning points, 
which limit the classical motion.  $U''$ is about 1\% bigger in the 
minimum on the right than in the minimum on the left.}
\label{fig2}
\end{figure}

If we restrict $\phi_{E,l}(s)$ and $\phi_{E,r}(s)$ to the discrete
values $s = n\hbar$ with integer $n$ between $-S$ and $S$, we obtain
the two vectors $\eta(E)$ and $\vartheta(E)$.  Our earlier discussion 
shows that, if $E$ happens to be an eigenvalue far enough from other 
eigenenergies, then $\phi_{E,l}$ or $\phi_{E,r}$ alone is a good 
approximation of the corresponding eigenvector.  Returning to the 
problem of resonance, we can find $E_1$ and $E_2$ by diagonalizing 
$\cal H$ in the subspace spanned by, say, $\phi_{E_1^{\rm (sc)},l}$ 
and $\phi_{E_2^{\rm (sc)},r}$.  Close to the semiclassical resonance 
(\ref{3}) there exist integers $m<0<n$ such that these functions are 
localized near $m\hbar$ and $n\hbar$, respectively.  We shall use the 
shorthand $\phi_m$ and $\phi_n$ for them.  Let $\chi$ denote the 
matrix of $\cal H$ so that $\chi_{ij}=\langle\phi_i|{\cal 
H}|\phi_j\rangle$, and let $o$ be the overlap matrix with elements 
$o_{ij}=\langle\phi_i|\phi_j\rangle$, where $i,j\in\{m,n\}$.  We 
recall that $o_{mn}$ is nonvanishing, although very small; we shall 
estimate it below.  There is no harm in supposing $\phi_m$ and 
$\phi_n$ to be normalized.

We now turn to an explicit calculation of the quantum-mechanical level
splitting. To a good approximation \cite{vhs4,vhs5,vhs6},
$E_1$ and $E_2$ agree with the eigenvalues $E_-$ and $E_+$ of 
the $2\times 2$ matrix
\begin{equation}
    o^{-1}\chi = {1\over 1-o_{mn}^2}\left( \begin{array}{cc} 
		 \chi_{mm}-\chi_{mn}o_{mn} & \chi_{mn}-\chi_{nn}o_{mn} \\
		 \chi_{mn}-\chi_{mm}o_{mn} & \chi_{nn}-\chi_{mn}o_{mn} \\ 
		 \end{array} \right)\ .
    \label{8a}
\end{equation}
To see it, expand the eigenfunctions in terms of $\phi_m$ and $\phi_n$
and take matrix elements. For the eigenvalues we find
\begin{equation}\label{eval}
       E_{\pm}={1\over 2(1-o_{mn}^2)}
       (\chi_{mm}+\chi_{nn}-2\chi_{mn}o_{mn}\pm D)
\end{equation}
where 
\begin{equation}\label{D}
       D=\left[(\chi_{mm}-\chi_{nn})^2+
       4(\chi_{mn}-\chi_{mm}o_{mn})
       (\chi_{mn}-\chi_{nn}o_{mn})\right]^{1/2}\ .
\end{equation}
Let $(x_\pm,y_\pm)$ denote the eigenvectors of $o^{-1}\chi$
corresponding 
to $E_\pm$. That is, the approximate eigenvectors of $\cal H$ are  
$x_\pm\phi_m+y_\pm\phi_n$. For $r_\pm = x_\pm/y_\pm$ an elementary 
computation yields
\begin{equation}\label{evec}
          r_\pm^2 = {\chi_{nn}-E_\pm\over \chi_{mm}-E_\pm}\ 
\end{equation}
and, by orthogonality, $r_+r_-=-1-(r_++r_-)o_{mn}$.  Near resonance 
Eqs.~(\ref{eval}) and (\ref{evec}) provide improved eigenvalues and 
eigenvectors as compared to the semiclassical ones.

We now have to tune $\delta$ to quantum-mechanical resonance.  Supposing
$\chi_{mn}$ is of the order of the overlap, the second term under the 
square root in (\ref{D}) is of order $o_{mn}^2$.  The overlap is a 
smooth function of $\delta$ and remains uniformly small in the small 
region where we vary $\delta$.  So the minimal distance between $E_+$ 
and $E_-$ is reached when $\Delta_{mn}=\chi_{mm}-\chi_{nn}$ vanishes 
(or is also of order $o_{mn}$).  Thus we conclude that the 
quantum-mechanical level splitting is of order $o_{mn}$.  
The order of 
magnitude of $o_{mn}$ is easily inferred from Eqs.~(\ref{wkbl}, 
\ref{wkbr}).  For this estimate we may suppose $E=E_1^{\rm (sc)}= 
E_2^{\rm (sc)}$.  Then
\begin{equation}\label{overlap}
o_{mn}=C \exp-{1\over \hbar}\int_{b_1}^{b_2}{\rm d}s\ {\rm arccosh}
\left({-E-F(s)\over\alpha a(s)}\right)
\end{equation}
where $C$ is a constant of order 1.
The exponential factor in $o_{mn}$ can be interpreted as a transition
probability. This suggests to write the level splitting $\Delta E=E_2-E_1$ 
in the form
\begin{equation}\label{split1}
\Delta E={\pi\hbar\over\tau_0}
         \exp-{1\over \hbar}\int_{b_1}^{b_2}{\rm d}s\ {\rm arccosh}
         \left({-E-F(s)\over\alpha a(s)}\right)
\end{equation}
where $1/\tau_0$ is an attempt frequency.
Recalling the expression of the 
hyperbolic cosine in terms of a natural logarithm and using 
$b_1\approx m\hbar$ and $b_2\approx n\hbar$, we find
the order-of-magnitude estimate
\begin{equation}\label{split2}
  \Delta E \approx {\pi\hbar\over\tau_0}
  \left[{\alpha S\over \Gamma(m^2+n^2)+\delta(m+n)}\right]^{n-m}\ .
\end{equation}
By identifying $\tau_0$ with the time period of the classical motion
 --~see Eq.~(\ref{9d}) below~--, Eqs.~(\ref{split1}) and (\ref{split2}) 
become fully explicit.  
For $\delta=0$ and $n=-m=S$ they agree, respectively,
with formulas (C.11) and (C.12) in Appendix~C of our earlier work \cite{vhs5}. 
In that case (\ref{split1}) leads to a remarkably
precise result, as can be seen in Table 1 of \cite{vhs5}. Equation 
(\ref{split1}) was obtained \cite{vhs6} also by an independent argument.

Of course $\delta$ can be chosen to let $\Delta_{mn}$
vanish. The reason is that $\chi_{mm}\approx E_1^{\rm (sc)}$ and
$\chi_{nn}\approx E_2^{\rm (sc)}$ (the small deviation coming from the 
fact that $\phi_m$ and $\phi_n$ do not satisfy the
boundary conditions (\ref{bc})), 
and {\em semiclassical} eigenvalues do cross each other
at resonance. Thus $\Delta_{mn}=0$ for a $\delta$ close to
$-(m+n)\Gamma$. In particular, because of a reflexion symmetry of the 
Hamiltonian, the level pairs $\{m=-n,n\}$ are at semiclassical {\em and} 
quantum-mechanical resonance, once $\delta=0$.

The origin of the level splitting is spin tunneling.  At 
quantum-mechanical resonance $|r_\pm|=1$ and thus the approximate 
eigenfunctions are $\psi_{\pm} = (\phi_{m} \pm \phi_{n})/\sqrt{2}$.  
If we start on the left, we take $\phi_{m} = (\psi_{+} + \psi_{-})/ 
\sqrt{2}$. It evolves under the influence of the dynamical
evolution generated by $\exp(it{\cal H}/\hbar)$.  After a time $T$ 
given by Eq.~(\ref{split1}) through $T\Delta E=\pi\hbar$,
the system is in $\phi_{n} = 
(\psi_{+} - \psi_{-})/\sqrt{2}$, i.e., on the right.  All this is 
exactly as in the case with a reflection symmetry $S_z \leftrightarrow 
-S_z$ such as when $\delta = 0$. In
the corresponding classical problem an approximate reflection symmetry 
survives for remarkably high values of $\delta$ ($\delta \le 
10\Gamma$) with a shifted center of symmetry; cf.  Figure~2.  Because 
of this approximate symmetry and for low enough energies, the attempt 
frequencies in the two, now different, orbits centered at $m$ and $n$ 
are hardly different; see Eq.~(\ref{9d}) below.  This fact is crucial 
for the interpretation of $\tau_0$ in the level splitting formula 
(\ref{split1}).

\section{Semiclassical Quantization}

Semiclassical quantization of a single spin can be handled
straightforwardly since we always find closed orbits -- if any -- as
the intersection of the energy surface ${\cal H} = -[\gamma S_z^2 +
\delta S_{z}] - \alpha S_{x} \equiv E$ and the sphere ${\bf S}^{2} =
S_{x}^{2} + S_{y}^{2} + S_{z}^{2} \equiv \sigma^{2}$; cf.  Fig.~1. 
Here the classical equations of motion associated with the
Hamiltonian ${\cal H} = -F(S_{z}) -\alpha S_{x}$ can be reduced to a
second-order differential equation \cite{vhs5.2} for $S_{z}$,
\begin{equation}
    \ddot{S}_{z} = -F(S_{z}) F'(S_{z})-E F'(S_{z})-\alpha^{2}
    S_{z} = -\left. {{\rm d} \over {\rm d} s}\, U(s) \,
    \right|_{s=S_{z}}
	\label{9a}
\end{equation}
where
\begin{equation}
	U(s) = {1 \over 2} F^{2}(s) + E F(s) + {1 \over 2} \alpha^{2}
	s^{2} \ .
	\label{9b}
\end{equation}
In the present case, $F(s) = \gamma s^{2} + \delta s$ so that
\begin{equation}
	U(s) = {1 \over 2} \gamma^2 s^4 + \gamma\delta s^{3} + [\gamma E
	+ {1 \over 2} (\alpha^{2} + \delta^{2})] s^{2} + \delta E s \ .
	\label{9c}
\end{equation}
Equation (\ref{9a}) describes the motion of a unit `mass' with
coordinate $s = S_{z}$ in a `potential' $U(s)$ so that its `energy'
${\cal E} = {1 \over 2} \dot{s}^{2} + U(s)$ is conserved.  The true
dimension of $\cal E$, which is called $\varepsilon$ in
Ref.~\cite{vhs5}, 
is (energy)$^{2}$.  In principle, we fix $\cal E$ by
specifying $s(0)$ and $\dot{s}(0)$.  In practice \cite[\S2.2]{vhs5}
there are only two independent constants of the three-dimensional
motion of a spin, viz., the energy $E$ and ${\bf S}^{2} = \sigma^{2}$,
so that $\cal E$ is bound to be a function of both of them: ${\cal E}
= {1 \over 2} (\alpha^{2} \sigma^{2} -E^{2})$.  This is most easily
verified by using (\ref{9b}), computing $E^{2}$, and realizing that $E
= -F(s) -\alpha S_{x}$ while $\dot{s} = - \alpha S_{y}$.

For $E$ negative enough, $U$ is a double-well potential, which is 
asymmetric in $s$ for $\delta \neq 0$ and we have two disjoint closed 
orbits, as is brought out by Figs.\ 1 and 2.  Figure 2 shows that the 
asymmetry develops much more slowly than the shift of the maximum; this
latter is roughly at $-\delta/2$.  The turning points 
$a_{1}<b_{1}<b_{2}<a_{2}$ are solutions of the equation $U(s)={\cal 
E}$.  Given $E$, classically allowed motion is between $a_{i}$ and 
$b_{i}$, with $i=1, 2$, so either on the left or on the right.  The 
period of this motion and, thus, implicitly the attempt frequency is
\begin{equation}
	T_{i}(E) = | 2 \int_{a_{i}}^{b_{i}} {\rm d}s \, 
	[2({\cal E}-U(s))]^{-1/2} | \ .
       \label{9d}
\end{equation}
If ${\cal E}$ is close enough to the bottom of the potential well, 
$U(s)$ is nearly parabolic in the domain of integration of (\ref{9d}) 
and $T_i(E)\approx 2\pi/ \sqrt{U''(s_i)}$, where $s_1$ and $s_2$ are 
the locations of the minima.  Because $U(s)$ is independent of the 
spin quantum number $S$, the attempt frequency will also be (nearly) 
independent of $S$.

Once an orbit exists, any $E$ is classically acceptable.  
Quantum-mechanically, however, only $2S+1$ energy eigenvalues survive.  
Determining the allowed eigenvalues to fair approximation and in 
closed form is what semiclassical quantization is (or should be) good 
for.

Handling a single spin, we have only a single pair of canonically
conjugate variables $q$ and $p$, which are related through the Poisson
bracket $\{q, p \} = 1$.  Since in a Hamiltonian formalism $q$ and $p$
are handled on an equal footing there is no harm in interchanging them
by putting $q_{\rm new} := -p$ and $p_{\rm new} := q$ so that the new
variables have the same Poisson bracket $\{ q_{\rm new}, p_{\rm new}
\} = 1$ as the old ones.  Instead of declaring $q = S_{z}$ and $p =
-\phi$ to be canonical coordinates \cite{vhs4,vhs5,vhs6}, with $\phi$
as the azimuth, we now find it advantageous to put
\begin{equation}
	q = \phi \qquad {\rm and} \qquad   p = S_{z} \ .
	\label{10}
\end{equation}
Semiclassical quantization is a condition on the action integral,
\begin{equation}
	\oint p \, {\rm d} q = \oint S_z \, {\rm d} \phi  = n h,
	\qquad \qquad n \in
	{{\bf Z}\hspace{-0.35em}{\rm Z}} \ ,
	\label{11}
\end{equation}
where the integral is to be taken over a classical, closed orbit, $h$
is Planck's constant, and $n$ is an integer; cf.\ Messiah \cite{m1}.
Along a closed orbit $S_{z}$ can often, certainly in the present case, 
be specified as a function of $0 \le \phi \le 2\pi$; cf.\ Fig.~1.
We write $S_{z} = \sigma \cos \theta$, with $\theta$ as the polar
angle, and arrive at the condition which we will use in the
spin problem below,
\begin{equation}
	\sigma\langle\cos\theta\rangle\equiv
	{\sigma \over 2\pi} \oint {\rm d}\phi \cos \theta = n \hbar,
	\qquad n \in {{\bf Z}\hspace{-0.35em}{\rm Z}} \ .
	\label{12}
\end{equation}

Of course $\cos \theta$ is to be given as a function of $\phi$.
The fact that $n$ is a positive {\em or negative} integer, restricted to 
$|n|\le S$, is typical to spins.

To verify that all this makes sense, we take the limit $\alpha \to 0$
so that the paraboloid $-[\gamma S_z^2 + \delta S_{z}] -\alpha S_{x}
\equiv E$ becomes very steep and, consequently, $S_{z}$ in (\ref{11})
is more or less constant as the spin tracks its orbit.  Hence we find
$ 2\pi S_{z} = n h$, which is equivalent to saying $S_{z} = n \hbar$,
as should be the case on the spectrum of $S_{z}$.  

As an application of the semiclassical quantization condition
(\ref{12}) we study the Hamiltonian (\ref{2}) with vanishing $\delta$.
The energy as given in polar coordinates,
\begin{equation}
	E = -\gamma S_{z}^{2} - \alpha S_{x} = - \gamma \sigma^{2}\cos^2
	\theta - \alpha \sigma \sin \theta \cos \phi \ ,
	\label{16}
\end{equation}
leads to a quadratic equation for $y:=x^{2}$ with $x = \cos \theta$;
this only happens when $\delta=0$.  Here we assume $E<0$, which is
typical to tunneling.  In view of considerations to come in \S4, we
introduce the dimensionless parameters 
\begin{equation}\label{parameters1}
	 a= {\alpha\over 2\gamma\sigma}\qquad
	 Q^{2}= -{E\over\gamma \sigma^{2}} \qquad \mbox{and}
	 \qquad \epsilon = {a\cos\phi\over Q^{2}}.
\end{equation} 
Here $a$ is supposed to be small; in the Mn$_{12}$ case, $a = 0.3$.

The quantity $y$ obeys the equation $y^{2} -2 Q^{2} (1-2 \epsilon^{2}
Q^{2}) y + (1-4 \epsilon^{2}) Q^{4} = 0$ so that
\begin{equation}
	y_{\pm} = Q^{2} \{ (1-2 \epsilon^{2} Q^{2}) \pm 2\epsilon \,
	[1 - Q^{2} + \epsilon^{2} Q^{4}]^{1/2} \} \ge 0 \ .
	\label{17}
\end{equation}
In view of $y_-(\phi)=y_+(\phi+\pi)$, it suffices to consider $y_+$ 
and, thus,
\begin{equation}
	x_{\pm} = \pm Q \{ (1-2 \epsilon^{2} Q^{2}) + 2\epsilon \,
	[1 - Q^{2} + \epsilon^{2} Q^{4}]^{1/2} \}^{1/2}
	\equiv \pm Q f(Q,\epsilon) \ .
	\label{18}
\end{equation}
That is, we are left with a `positive' branch, $x_{+}(\phi)$, and a 
`negative' one, $x_{-}(\phi)$, symmetrically positioned with respect to
the $S_{x}$-$S_{y}$ plane.  Both $x_+$ and $x_-$ have to be inserted
into 
(\ref{12}).  They correspond to $n>0$ and $n<0$, respectively, and yield 
the same energy.  Indeed, (\ref{12}) and (\ref{18}) imply
\begin{equation}\label{Eeven}
	 \left({n\over S}\right)^2=Q^2 \langle f\rangle^2\ .
\end{equation}
As a matter of fact, $f$ is a function of $Q^2$ so that the solution of 
(\ref{Eeven}) for $E$ only depends on $|n|$.  It then remains to
calculate the quantum-mechanical splitting of the levels $-n$ and $n$
as indicated in \S2.

In the present case, semiclassical quantization is a straightforward
integration giving up to second order in $a$ (see Appendix A)
\begin{eqnarray}
	\langle \cos \theta \rangle &=& {1 \over 2\pi} \int_{0}^{2\pi}
	{\rm d} \phi \,\, x_{\pm}(\phi) = \nonumber \\
	&=& \pm Q [1 - {a^{2} \over 4} (Q^{-2} + Q^{-4})] =
	{n \hbar \over \sigma} =  {n \over S} \ .
	\label{18a}
\end{eqnarray}
The dependence of $ x_{\pm}(\phi)$ on $\phi$ is a dependence upon 
$\cos \phi$.  If desired, one can change variables through $z:= \exp 
(i \phi)$ and obtain a contour integral in the complex $z$ plane.

What we are after is the energy $E$ as it appears in $Q$.  First we 
solve (\ref{18a}), a fourth order equation in $Q$.  Recalling that it 
already contains an error term of the order of $a^4$, it suffices to 
find $Q=Q(a)$ up to second order, which can be done by iteration.  To 
this end we rewrite (\ref{18a}) in a form that is easy to iterate,
\begin{equation}
	Q = {|n| \over S} + {a^{2} \over 4} (Q^{-1} + Q^{-3}) \ .
	\label{19}
\end{equation}
Iterating once, i.e., replacing $Q$ in the right-hand side of (\ref{19})
by $|n|/S$, we obtain $Q$ up to second order in $a$:
\begin{equation}
	Q = {|n| \over S} \left\{ 1 + {a^{2} \over 4}
	\left( {S \over n} \right)^{2} \left[ 1 + \left(
	{S \over n} \right)^{2} \right] \right\} \ .
	\label{19a}
\end{equation}
Because $E= -\gamma\sigma^{2} Q^{2}$, squaring (\ref{19a}) and 
dropping terms of order $a^4$ we obtain the second-order expression 
for the frequencies ($|n| \le S$),
\begin{equation}
	\Omega_{n} \equiv  E_{n} / \hbar = - n^{2} \Gamma - {\alpha^{2}
	\over 8 \Gamma} \left[ 1+ \left( {S \over n} \right)^{2}
\right]\ .
	\label{20}
\end{equation}
Here and in the next section we have dropped the superscript `sc', and
$E_n$ denotes a semiclassical eigenvalue.
As announced, the degeneracy of $E_{n}$ and $E_{-n}$ is not lifted.  
It has to be borne in mind that (\ref{20}) has been obtained under the 
assumption that we may drop everything beyond second order in $a^2$, 
an assumption that may, but need not, hold.  In Appendix A we derive 
an exact expression of $\langle\cos\theta\rangle$ in terms of a power 
series in $a$.  Using this power series one can obtain semiclassical 
energies up to any order.

Equation (\ref{20}) being second-order in $a$, it is instructive to 
compare it with second order quantum-mechanical perturbation theory 
\cite{jrf}, 
\begin{equation}
	E_n^{(2)}/\hbar = -n^2 \Gamma
	-\frac{\alpha^2}{8\Gamma}
	\left[\frac{n^2 + S(S+1)}{n^2-1/4}\right]\ .
	\label{20a}
\end{equation}
The correction to $-n^2\Gamma$ is slightly bigger in absolute value 
than in (\ref{20}) but the agreement is excellent, except for, say, 
$|n| \le 2$ and $S \le 4$.  For the ground state with $|n| = S$, 
Eq.~(\ref{20}) yields $E_{\pm S}=-\gamma\sigma^2-\alpha^2 /4\gamma$ 
which agrees with the minimal classical energy, cf.  
\cite[Eq.~(2.6)]{vhs5}.  Table I of Ref.~\cite{vhs5} shows that, for 
$\alpha = \gamma = \hbar = 1$ and $S \ge 8$, the ground state has 
$\Omega_{S} + S^{2} \Gamma = -0.26$, which is indeed near the 
predicted $-1/4$.  In fact, it is slightly less, as we would expect.  
Under the proviso $S \ge 5$ and $|n| \ge 3$, the deviation of the 
`shift' $E_n^{(2)}/\hbar + n^2 \Gamma$ from that given by numerically 
exact eigenvalues $E_n$ is less than $15$\%.

\section{Determining the Energies for Nonzero $\delta$}

We proceed in analogy to the $\delta=0$ case.  In the argument below 
we are looking for solutions of a fourth-order equation $P_4(x)=0$ in 
dependence upon a given combination $a$ of the coefficients of the 
polynomial $P_4$.  Instead of attempting to obtain an exact solution, 
which would not provide much insight, we rewrite $P_4(x)=0$ in the 
form of a fixed-point equation $x=f(x,a)$, and, supposing the 
smallness of $a$, find the solution $x=x(a)$ by iteration, up to a 
given order in $a$.  We have already applied this procedure once, 
viz., to (\ref{19}).

The energy as given in polar coordinates and to be compared with
(\ref{16}) 
reads
\begin{equation}
	E = -\gamma S_{z}^{2} -\delta S_z -\alpha S_{x} =
	- \gamma \sigma^{2} \cos^2 \theta - \delta \sigma \cos \theta
	-\alpha \sigma \sin \theta \cos \phi \ ,
	\label{21}
\end{equation}
As before, we put $x = \cos \theta$ but do not get a quadratic equation
in $y:=x^{2}$ once $\delta \neq 0$,
\begin{equation}
	\gamma \sigma^2\,x^2 + \delta \sigma\,x + \alpha \sigma
	\sqrt{1-x^2}\,\cos \phi + E = 0 \ .
	\label{22}
\end{equation}
In agreement with (\ref{parameters1}) we now define the dimensionless
quantities
\begin{equation}\label{parameters2}
	a={\alpha\over 2\gamma\sigma}\qquad 
	d={\delta\over 2\gamma\sigma}\qquad
	Q^2=-{E\over\gamma\sigma^2}+d^2\qquad
	\epsilon={a\cos\phi\over Q^2}\ 
\end{equation}
which reduce to (\ref{parameters1}) whenever $\delta=0$.  
Here too, $E<0$ will be assumed. 

The two solutions $x=x_\pm$ of (\ref{22}) obey the equation
\begin{equation}
	x = -d \pm Q \left(1-2 \epsilon
	\sqrt{1-x^2} \right)^{1/2} \ .
	\label{24}
\end{equation}
According to what has been outlined at the beginning of this section, 
we assume $\epsilon<1/2$ so that the outer square root can be 
expanded.  Then (\ref{24}) can be solved by iteration to any order in 
$a$.  The algebra has been relegated to Appendix B. After an 
integration with respect to $\phi$ one finds, to second order in $a$,
\begin{equation}
	\langle x \rangle \equiv {1 \over 2\pi} \int_{0}^{2\pi}
	{\rm d}\phi\,\, x(\phi) = x_{0}+\frac{a^{2}}{2Q^{4}}\,[x]_{2}
	\label{25}
\end{equation}
where $x_{0} = -d \pm Q$ and
\begin{equation}
  [x]_{2} = \mp Q \,[ \pm Q x_{0} + {1 \over 2}(1-x_{0}^{2})]
  = \mp {Q \over 2} \,(Q^{2} +1 -d^{2}) \ .
	\label{26}
\end{equation}
Unless stated otherwise, we will not repeat that henceforth we have
to add a term ${\cal O} (a^{4})$ to all right-hand sides of the
equations in this section.

Semiclassical quantization means, in complete analogy to (\ref{18a}),
that $\langle x \rangle$ is to be equal to $n/S$ with $-S \le n \le
S$.  Realizing that $n>0$ corresponds to the upper and $n<0$ to the
lower sign in (\ref{26}) and $\pm n = |n|$, we combine (\ref{26}) with
(\ref{25}) and obtain
\begin{equation}
	{n \over S} = \langle x \rangle = \pm Q -d \mp {a^{2} \over 4Q}
	[1 + Q^{-2}(1-d^{2})]
	\label{27}
\end{equation}
and thus
\begin{equation}
	{\pm n \over S} = {|n| \over S} = Q - {\rm sgn}(n)\, d -
	{a^{2} \over 4Q} [1 + Q^{-2}(1-d^{2})] \ .
	\label{28}
\end{equation}
We now rewrite this in a form that is apt to iteration,
\begin{equation}
	Q = Q_0 +
	{a^{2} \over 4Q} [1 + Q^{-2}(1-d^{2})] \ ,\qquad 
	Q_0={|n|\over S}+{\rm sgn}(n)\, d \ .
	\label{29}
\end{equation}
To obtain $Q$ to second order in $a$, we iterate once,
\begin{equation}
	Q = Q_0 +
	{a^{2} \over 4Q_0} [1 + Q_0^{-2}
	(1-d^{2})]  \ .
	\label{30}
\end{equation}
For $\delta = d = 0$ we recover (\ref{19a}).  Squaring (\ref{30}),
dropping terms of order $a^4$ and taking advantage of
(\ref{parameters2}) we
obtain
\begin{equation}
	 \Omega_{n} \equiv E_n/\hbar=
	 -n^{2}\Gamma -n \delta - {\alpha^{2} \over 8\Gamma} 
	 \left[{S^2+n^2+n\delta/\Gamma\over (n+\delta/2\Gamma)^2}
	 \right]  +
	 {\cal O}(a^{4}\Gamma S^{2}) \ .
	\label{31}
\end{equation}
Plainly, this is identical with (\ref{20}) for $\delta = d = 0$.
On the other hand, second-order perturbation theory gives
\begin{equation}
	E_{n}^{(2)}/\hbar = - n^{2} \Gamma -n \delta -{\alpha^{2}
	\over 8\Gamma} \left[\frac{S(S+1) + n^{2}+ n \delta/\Gamma}
	{(n + \delta/2\Gamma)^{2}-1/4} \right] \ ;
	\label{32}
\end{equation}
see also \cite{jrf}.
The above expression agrees with Eq.~(\ref{20a}) when $\delta$ 
vanishes, and with (\ref{31}) whenever $n$ and $S$ are sufficiently 
large and $a\ll 1$.  The latter condition is quite reasonable since we 
have used quantum-mechanical perturbation theory and, thus, compared 
the `perturbation' $-\alpha S_{x}$ with the `rest', viz., $-\gamma 
S_{z}^{2}-\delta S_z$.

Once we know the semiclassical energies $E_{n}$, we can tune $\delta$ 
so as to get semiclassical resonance $E_{m} = E_{n}$ for some $m \neq
n$.  
Despite being obtained for $\alpha=0$, the resonance condition (\ref{3}) 
remains valid for $\alpha\neq 0$ as well.  Both (\ref{31}) and 
(\ref{32}) depend on $n$ through the expression $n^{2}+ n 
\delta/\Gamma$.  Therefore (\ref{3}) directly 
implies $E_m=E_n$ for each couple $\{m,n\}$ satisfying 
$m+n=-k=-\delta/\Gamma$.  

A closer inspection shows that (\ref{3}) implies degeneracy, to
\emph{any} order, of semiclassical eigenvalues.  To see
why, we return to Eq.~(\ref{24}).  Let us start by averaging it, viz.,
\begin{equation}
	{n\over S}=-d\pm
Q\langle(1-2\epsilon\sqrt{1-x^2})^{1/2}\rangle\       
	\ .
	\label{32a}
\end{equation}
We now add $d$ on the right and on the left, multiply both sides by
$S$, and square the result so as to find
\begin{equation}
	n^2 + n\delta/\Gamma = -(\delta/2\Gamma)^2 + 
        \left[SQ \langle (1-2\epsilon\sqrt{1-x^2})^{1/2}\rangle\right]^2 \ .
	\label{32b}
\end{equation}
The right-hand side does not show any $n\,$-dependence. It is obtained
by
taking $x$ as a solution to the fourth-order equation (\ref{24}), 
depending on $E,\,\alpha,\,\delta,\,\Gamma$, and $\phi$.  After 
integration with respect to $\phi$, we are left with (\ref{32b}), an 
implicit equation for $E$.  Solving it for $E$, the solution $E=E_{n}$ 
will depend on $n$ through the combination $n^2 + n\delta/\Gamma$, as 
it shows up in (\ref{32b}).  Hence $E_{m} = E_{n}$ whenever $m^2 + 
m\delta/\Gamma=n^2 + n\delta/\Gamma$.

\section{Extensions}

The extension of the above results to half-integer spins is 
straightforward.  All that we have to do is interpreting $m$ and $n$
as half-integers whenever they refer to eigenvalues of $S_z$.  In 
particular, $m$ and $n$ are half-integers in the resonance condition 
(\ref{3},\ref{4},\ref{5}) and in the quantization condition 
(\ref{11},\ref{12}), and the semiclassical eigenvalues are also 
labelled by half-integers.  At semiclassical resonance $\delta/\Gamma$ 
is still an integer.

The case where in the Hamiltonian $S_x$ is replaced by $S_x^l$ with
$l>1$
brings nothing new, if $l$ is an odd integer. For positive integer $l$
semiclassical quantization is essentially unchanged. In the definition 
(\ref{parameters2}) of $\epsilon$, $\cos\phi$ is replaced by
$\cos^l\phi$
and in Eqs.~(\ref{24}), (\ref{32a}) and (\ref{32b}) $(\sqrt{1-x^2})^l$
is substituted for $\sqrt{1-x^2}$. The upshot is that the semiclassical
eigenvalues still depend on $n$ through the expression
$n^2+n\delta/\Gamma$
and therefore (\ref{3}) still implies $E_m^{\rm (sc)}=E_n^{\rm (sc)}$.
If $l$ is odd, spin tunneling splits the semiclassical degeneracy.

For even $l$, however, the meaning of Eqs.~(\ref{3})-(\ref{5}) changes.
Depending on the parity of $k$, (\ref{5}) may refer to resonance
or to true degeneracy. Let us recall that (\ref{3}) gives the
condition for degeneracy when $\alpha=0$. In the case of integer spins,
if $k$ is odd then $n-m=-k-2m$ is odd for each $m$ and $n$ such that
$m+n=-k$.
As a consequence, the $\alpha S_x^l$ term does not split the degeneracy
of all these level pairs. In the case of half-integer spins the same
happens if $k$ is even. Therefore 
$E_m=E_n$ holds for the true eigenvalues, even if not {\em exactly} 
at, but only very close to, the field $\delta=k\Gamma$ at which 
$E_m^{\rm (sc)}=E_n^{\rm (sc)}$. 

An explanation of this generalized Kramers degeneracy can 
be found in \S8 of \cite{vhs6}, but for the reader's convenience we
repeat the argument here. We write the matrix of $\cal H$ (for $l$ even)
in the basis of the $S_z$-eigenvectors $|n\rangle$. 
By noticing that $\langle m|{\cal H}|n\rangle=0$ whenever $n-m$ is odd,
we can permute the elements of the basis so as to
transform the matrix into a block-diagonal form containing two blocks:
the first is formed with $n=-S,-S+2,-S+4,\ldots$, the second with 
$n=-S+1,-S+3,-S+5,\ldots$. Both can be diagonalized independently, 
so the von Neumann-Wigner argument \cite{nw} about the
avoided level crossing does not apply.

For a degenerate energy level there is no tunneling. That is to say,
a left-localized initial state remains approximately left-localized
during time evolution. The reason is that in the corresponding 
two-dimensional eigensubspace of the Hamiltonian there exists an
approximately
left-localized state which is dominant in the eigenfunction
expansion of the initial state. Therefore, if in an experimental 
situation the tranverse anisotropy fields are a combination of 
even powers only, measurements {\em \`a la} \cite{Mna,Mnb} of the
quantum 
decay of magnetization in an increasing longitudinal external magnetic 
field would ideally yield a fast decay (a step), only when $\delta/
\Gamma$ passes every other integer value. In other words,
every second (expected) step would be missing. Unfortunately, 
an experiment in an applied magnetic field usually has a transverse 
component that induces steps at the remaining integer values.

\section{Discussion}

We have defined semiclassical resonance to be the degeneracy $E_{m} = 
E_{n}$ for energy levels as they follow from semiclassical 
quantization.  The ensuing level splitting is a purely 
quantum-mechanical phenomenon.  Both can be handled straightforwardly 
by the WKB formalism \cite{vhs4,vhs5,vhs6}.  The resonance experiments 
as set up for spin tunneling in ${\rm Mn}_{12}$ acetate 
\cite{NoSe}--\cite{Mnd} and Fe$_{8}$ magnetic crystals \cite{sangreg}
are both elegant and quite promising since they provide a detailed 
test of the physics of spin quantum tunneling and the crossover from 
the regime of thermal activation to that of quantum behavior.  The 
present paper's considerations on tunneling resonance are valid, if 
the temperature is low enough; say \cite{cg,gc} below $1^{^\circ}$K in 
the ${\rm Mn}_{12}$ case.

If so, one could apply semiclassical quantization (\S3) to arbitrary 
numerical precision but the amount of insight thus obtained is fairly 
restricted.  We have chosen a different way out and derived an 
explicit expression for the energy levels and, hence, for 
semiclassical resonance through an external field $-\delta S_{z}$ 
under the assumption that $\alpha / (2 \Gamma S Q^{2}) \ll 1$ where 
$Q^{2} = (\delta^{2}/4 - \gamma E)/(\Gamma S)^{2}$ is dimensionless; 
cf.\ (\ref{parameters2}).  For the ground state of ${\rm Mn}_{12}$ 
with $S= 10$, we have $Q \approx 1$ and the considerations apply.  In 
fact, we have also derived the energy levels for $\delta = 0$ under 
slightly less restrictive conditions and found that the shift of the 
ground state energy is well-predicted by $-\alpha^{2} / 4\Gamma$.  In 
some resonance experiments one would like to focus on states with $m$ 
and $n$ closer to zero since, in resonance, the tunneling frequency is 
much higher.  As a consequence, the afore mentioned assumption may, 
but need not, hold.  It mainly depends on $Q^{2}$.  If $E \approx 
-n^{2} \hbar \Gamma$, then $Q^2 \approx (n/S)^{2}$ and it is the 
smaller the closer $n$ is to $0$.

Whenever $\delta$ is tuned to degeneracy of some energy levels of the 
Hamiltonian ${\cal H}_0 = -\gamma S_z^2 - \delta S_z$, this degeneracy 
is not lifted by semiclassical quantization of the system with 
$\alpha\neq0$.  Only quantum mechanics does lift it, and only in high 
orders of $\alpha /\Gamma S$.  At a first sight rather surprisingly -- 
but at a first sight only (\S2) -- the mathematics underlying 
tunneling for $\delta\neq 0$, when the Hamiltonian's symmetry of a 
rotation through $\pi$ about the $S_{x}$ axis is broken, is identical 
to that of the case $\delta = 0$.  That is to say, the very same 
formalism, whether WKB or otherwise, can be applied to both cases.

\subsection*{Acknowledgments}

The authors thank the Max Planck Institute for the Physics of Complex
Systems for its hospitality and support during their stay in Dresden,
where most of this work was done.  A.  S\"{u}t\H{o} is greatly indebted
to the Hungarian Scientific Research Fund (OTKA) for additional
support under grant No.~T14855.

\section*{Appendix A: Semiclassical Quantization for $\delta =0$}

We have to integrate $x=\cos\theta$ relative to $\phi$, i.e., 
average, the expression $\{ \ldots \}^{1/2}$ in (\ref{18}) and below,
\begin{equation}
	x_{\pm} = \pm Q \{ 1-2 \epsilon^{2} Q^{2} + 2\epsilon \,
	[1 - Q^{2} + \epsilon^{2} Q^{4}]^{1/2} \}^{1/2} \ ,
	\label{a0}
\end{equation}
where $\epsilon = Q^{-2} a \cos \phi$.  For $\delta = 0$, there are
two disjoint classical orbits once $E < -\alpha \sigma$.  Exactly in
this energy range one finds $\{ \ldots \} \ge 0$ so that taking the
square root in (\ref{a0}) is well-defined.  For the moment we suppose
that $a$ and thus $\epsilon $ is small enough to
perform a series expansion of the outer square root. It is of the form
\begin{equation}
       (1-2y)^{1/2} = 1-\sum_{n \ge 1} b_{n} y^{n} \quad {\rm where}
       \quad b_n = {(2n-3)!! \over n!} \ .
	\label{a1}
\end{equation}
The series converges for $|y|<1/2$ while $b_1=1,\, b_2=b_3=1/2$. 
Introducing the short-hand $r = [1 - Q^{2} + (\epsilon Q^2)^{2}]^{1/2} $
we see that we have to average
\begin{equation}
	\{ 1 -2\epsilon (\epsilon Q^{2} - r) \}^{1/2} = 1 - \sum_{n\ge1}
	b_{n} \epsilon^n (\epsilon Q^2 -r)^{n} \ .
	\label{a2}
\end{equation}
We now concentrate on the series, insert the definition of $\epsilon$ 
and apply the binomial theorem to (\ref{a2}) so as to find
\begin{equation}
	\sum_{n \ge 1}  \sum_{m=0}^{n} \, b_{n} Q^{-2n} a^{m+n}
(-1)^{n-m}
	{n \choose m} r^{n-m} \cos^{m+n} \phi \ .
	\label{a3}
\end{equation}
At this point we observe that $r$ depends on $\phi\,$ through 
$\cos^2\phi$.  Hence, $\langle g(r) \cos^{\ell} \phi \rangle$ vanishes 
for $\ell$ odd, whatever the function $g$ may be.  Thus, when 
averaging (\ref{a3}), only terms with $\ell = m+n$ even will survive.  
For these terms $n-m=(n+m)-2m$ is also even and we can apply the 
binomial theorem to $r^{n-m}$ as well.  Furthermore, averaging a 
cosine is equivalent to a contour integration over the unit circle 
C$_{1}$ in the complex $z$ plane,
\begin{equation}
	\langle \cos^{\ell} \phi \rangle = {2^{-\ell} \over 2\pi i}
	\oint_{{\rm C}_{1}} {{\rm d}z \over z} \, (z+z^{-1})^{\ell} =
    2^{-\ell} {\ell \choose \ell/2} \ .
	\label{a4}
\end{equation}
Finally, we are left with
\begin{eqnarray}
    \langle\cos\theta\rangle=\pm Q\left[1-
	\sum_{n \ge 1}  \sum_{m=0}^{n} \sum_{l=0}^{(n-m)/2} \,
	b_{n} Q^{-2n} (1-Q^2)^{{n-m\over 2}-l} 
	\left( {a \over 2} \right)^{2l+m+n}\right. \nonumber\\
       \left. {n \choose m} {{1 \over 2}(n-m) \choose l}
	{2l+m+n \choose l + {1 \over 2} (m+n)} \right]
	\equiv \pm Q\left[1-\sum_{k=1}^\infty {\cal C}_k(Q^2)\,
	\left({a\over 2}\right)^{2k}\right]
	\label{a5}
\end{eqnarray}
where it is understood that $m+n$ is even.  If it were not because of 
the third binomial coefficient in (\ref{a5}) we could directly resum 
the series.  In the style of Borel \cite{titch,hardy}, however, we 
could, by noting $n!  = \Gamma(n+1)$ and taking an Eulerian integral 
representation of the second kind \cite{c,ww} for the Gamma function 
in conjunction with Hankel's contour integral \cite{c,ww} for its 
inverse,
\begin{equation}
	{1 \over \Gamma(t)}= {i \over 2\pi}\int_{{\rm C}_{2}}
	{{\rm d}z \over z} \, e^{-z}\,(-z)^{-t} \ ,     
	\label{a5bis}
\end{equation}
where ${\rm C}_{2}$ is a counter-clockwise infinite contour around the 
positive real axis.  We will not pursue this idea here.  Collecting 
terms of order $a^{2}$ we find ${\cal C}_1=Q^{-2}+Q^{-4}$ and arrive 
at (\ref{18a}), as advertised.

How good is all this?  Since we interchange the series (\ref{a1}) and 
averaging, viz., $\oint {\rm d}\,z \ldots$, we have to prove uniform 
convergence of the series and thus, in view of (\ref{a1}), uniform 
boundedness of $y=\epsilon (\epsilon Q^{2} - r)$ as $\cos \phi$ varies 
between 1 and $-1$.  The upshot appears in (\ref{a5}).  Plainly, the 
series in (\ref{a5}) is convergent for $a$ small enough so that, for a 
given $a$, $Q$ and, thus, the energy has to stay away from zero.  The 
series in (\ref{a5}) certainly diverges at energies for which the two 
classical orbits merge into a single one.  One can handle divergent 
series through, e.g., Borel summation \cite{titch,hardy} as above, but 
this is not the topic of the present paper.

\section*{Appendix B: Quantizing to Second Order in $\epsilon$}

In this appendix we consider $\epsilon$ as a `small' parameter and
sketch how one can perform semiclassical quantization up to second
order in $\epsilon$. As we will see, the method can be generalized
and the considerations of Appendix A concerning the validity of the
approximations involved apply here as well.

We denote the two branches $x_{\pm}$ of (\ref{24}), viz.,
\begin{equation}
	x = -d \pm Q [1-2\epsilon \sqrt{1-x^{2}}]^{1/2} \ ,
	\label{b1}
\end{equation}
simply by $x$.  Keeping in mind that $x=\cos \theta$ depends on 
$\phi$, we fix $\phi$ for the moment and study $x$'s dependence upon 
$\epsilon = aQ^{-2} \cos \phi$.  In view of (\ref{b1}) it is clear 
that $x=x(\epsilon)$ is analytic in a neighborhood of $\epsilon = 0$.  
Defining the $n$th partial `Taylor sum' of $x(\epsilon)$ to be
\begin{equation}
	x_{n} = \sum_{k=0}^{n} {x^{(k)}(0) \over k!}\,\epsilon^{k}
\equiv
	\sum_{k=0}^{n} \, [x]_{k}\,\epsilon^{k} \ ,
	\label{b2}
\end{equation}
with $x_{0} = x(0) = -d \pm Q$, using (\ref{a1}), and expanding the 
outer square root of (\ref{b1}), we can rewrite (\ref{b1}) in the form
\begin{equation}
	x = x_{0} \mp Q \sum_{n \ge 1} b_{n} \epsilon^{n}
	(1-x^{2})^{n/2} \ .
	\label{b3}
\end{equation}
Up to and including terms of order two we then find
\begin{equation}
	x = x_{0} \mp Q [ \epsilon (1-x^{2})^{1/2} + 
	\epsilon^{2} (1-x^{2})/2] + {\cal O}(\epsilon^{3}) \ ,
	\label{b4}
\end{equation}
whence
\begin{equation}
	x_{1} = x_{0} \mp Q (1-x_{0}^{2})^{1/2}\, \epsilon
	= x_{0} + [x]_{1} \epsilon
	\label{b5}
\end{equation}
because $\sqrt{1-x^{2}} = \sqrt{1-x_{0}^{2}} +{\cal O}(\epsilon)$.  

In order to obtain $x_2$, we observe that (\ref{b4}) implies
\begin{equation}\label{[x]_2}
   [x]_2=\mp Q\{[\sqrt{1-x^2}]_1 + [1-x^2]_0\ /2 \} \ .
\end{equation}
In analogy to (\ref{b2}), we have used the notation 
$[f]_n=f^{(n)}(0)/n!$.  

As for the entries of (\ref{[x]_2}), a simple calculation gives
\begin{eqnarray}
	[\sqrt{1-x^{2}}]_{1} &=& \left.{{\rm d} \sqrt{1-x^{2}} \over
	{\rm d}\epsilon}\,\,\right|_{\epsilon = 0} =
	-x_{0}(1-x_{0}^{2})^{-1/2}x'(0) \nonumber\\
	&=& -x_{0}(1-x_{0}^{2})^{-1/2}[x]_{1} = \pm Q x_{0} \ .
	\label{b7}
\end{eqnarray}
Since $[1-x^{2}]_0 = 1-x_{0}^{2} $ we end up with
\begin{equation}
	x_{2} = x_1+[x]_2\epsilon^2=
	x_0\mp Q\epsilon\sqrt{1-x_0^2} - Q \epsilon^{2} [ Qx_0 \pm
	(1-x_0^{2})/2 ]
	\label{b8}
\end{equation}
It is $x_{2}$ that is to be averaged with respect to $\phi$.  Inserting 
$\langle \epsilon \rangle = 0$ and $\langle \epsilon^{2} \rangle = 
a^{2}/2Q^{4}$ into $\langle x_2\rangle$ we obtain Eq.~(\ref{25}).

\end{document}